# STATUS OF THE SEARCH FOR SUPERSYMMETRIC DARK MATTER

*David. B. Cline**
*Astrophysics Division*
*Department of Physics & Astronomy*
*University of California, Los Angeles, CA 90095 USA*

**Abstract.** We assume the supersymmetric model for dark matter in the universe and our galaxy, and direct methods to distinguish these kinds of dark matter are described. We then focus on the current and future experiment search for SUSY-WIMPS. Theoretical models suggest that a new generation of at least one ton detectors may be required to observe this form of dark matter. We concentrate on Liquid Xenon detectors because they can be scaled to large mass.

## 1. SUSY Models for Dark Matter Detection and Expectations

The search for dark matter particles is among the most fundamental of all astroparticle physics goals. The first evidence for dark matter came in 1933 and confirmation came in 2003 from the WMAP data (see Table 1). We know that at least 80% of the matter in the universe is due to this source. An excellent guide to the search is given by the SUSY-WIMP model [1]. As new collider physics results have appeared the calculations for the rate of such WIMPs in dark matter detectors has gone down [2]. Currently the expected value is less than $10^{-2}$ events/kg/day.

The search for SUSY WIMPs requires very powerful dark matter detectors that can be scaled to one ton if necessary. The leading methods use cryogenic detectors (CDMS, Edelweiss, CREST, etc.) and Liquid Zenon (ZEPLIN I, II, IV, etc., XENON, XMASS, etc.) [1].

The detection of SUSY WIMPs at this level could require a massive (one ton or even greater mass) powerful discriminating (ZEPLIN II) detector that we call ZEPLIN IV. The ZEPLIN II detector is under construction at UCLA and elsewhere for operation at Boulby in 2002. We consider this a prototype for ZEPLIN IV. CDMS II is about to start operation at SOUDAN and other detectors are about to turn on. Recently the new results for the (g-2) of the muon and other new calculation suggest that supersymmetry may be a correct model [2]. Also the expected mass of the LSP seems to be larger, requiring more powerful dark matter detectors.

## 2. Current Search Results: DAMA/ZEPLIN/Edelweiss/CDMS I, etc.

The current status of the search for WIMPs is somewhat confused; a recent summary can be found in Ref. 1:

1. The sensitivity level is about 1/5 event/kg/day [1];
2. The DAMA group is making claims for a discovery; by observation of a possible annual signal variation [1];
3. The CDMS group has shown that their data and those of DAMA are incompatible to 99.5% confidence level against the observation of WIMPs [1].

4. The Edelweiss and ZEPLIN I groups have excluded the mean value of the DAMA results [1].

5. Recent ZEPLIN I results give the best current limits, well below the DAMA results [3].

## 3. Next Step in the Search for SUSY WIMPs

These recent results barely reach the level expected for SUSY WIMPs [1][2]. In Table 2 we show some current detector information taken from my recent article in Scientific American (March 2003).

---

**TABLE 1. \*\***

Dark Matter

In ~1933 ⇒

> **W Map
> 2003
> Definite
> Evidence for
> Dark Matter**

F. Zwicky showed that Galactic Clusters must have
"missing mass" due to the high velocities of the galaxies.

---

** C.L. Bennet et al, Astro-PH 0302207.

# Leading Searches for Dark Matter

| Project | Location | Start Date | Primary Detector Type | Primary Detector Material | Primary Detector Mass (kg) | Discrimination Detector Types(s) |
|---------|----------|-----------|----------------------|--------------------------|---------------------------|----------------------------------|
| UKDMC | Boulby, UK | 1997 | Scintillation | Sodium iodide | 5 | None |
| DAMA | Gran Sasso, Italy | 1998 | Scintillation | Sodium iodide | 100 | None |
| ROSEBUD | Canfranc, Spain | 1999 | Cryogenic | Aluminum oxide | 0.05 | Thermal |
| PICASSO | Sudbury, Canada | 2000 | Liquid droplets | Freon | 0.001 | None |
| SIMPLE | Ristrel, France | 2001 | Liquid droplets | Freon | 0.001 | None |
| DRIFT | Boulby, UK | 2001 | Ionization | Carbon disulfide gas | 0.16 | Directional |
| Edelweiss | Frejus, France | 2001 | Cryogenic | Germanium | 1.3 | Ionization, thermal |
| ZEPLIN I | Boulby, UK | 2001 | Scintillation | Liquid Xenon | 4 | Timing |
| CDMS II | Soudan, Minn., US | 2003 | Cryogenic | Silicon, germanium | 7 | Ionization, thermal |
| ZEPLIN II | Boulby, UK | 2003 | Scintillation | Liquid Xenon | 30 | Ionization, scintillation |
| CRESST II | Gran Sasso, Italy | 2004 | Cryogenic | Calcium tungsten oxide | 10 | Scintillation, thermal |

## (a) ZEPLIN II at Boulby

Starting in the early 1990s, the UCLA/Torino ICARUS group initiated the study of liquid Xe as a WIMP detector with powerful discrimination [4][5].  The basic mechanism of detection in Liquid Xenon is shown in Fig. 2. Our most recent effort is the development of the two-phase detector.  Figure 3 shows our 1-kg, two-phase detector and the principle of its operation.  WIMP interactions are clearly discriminated from all important background by the amount of free electrons that are drifted out of the detector into the gas phase where amplification occurs.  In Fig. 4, we show the resulting separation between backgrounds and simulated WIMP interactions (by neutron interaction) [4].  It is obvious from this plot that the discrimination is very powerful.

Construction has begun on a large two-phase detector to search for WIMPs [6].  The UCLA/Torino group has formed collaboration with the UK Dark Matter team to construct a 40-kg detector (ZEPLIN II) for the Boulby Mine underground laboratory (Fig. 5) [6]. Table 3 lists the important properties of Liquid Xenon.

Continuation of the R&D effort with liquid Xe to attempt to amplify the very weak WIMP signal [6]. The second idea to test is inserting a CsI internal photo cathode to convert UV photons to electrons that are subsequently amplified by the gas phase of the detector [7]. In Fig. 5 we show the latest design of the ZEPLIN II detector. The expected sensitivity of ZEPLIN II is shown in Figure 8. This detector is being assembled at Rutherford Laboratory now for installation in the Boulby underground laboratory later this year.

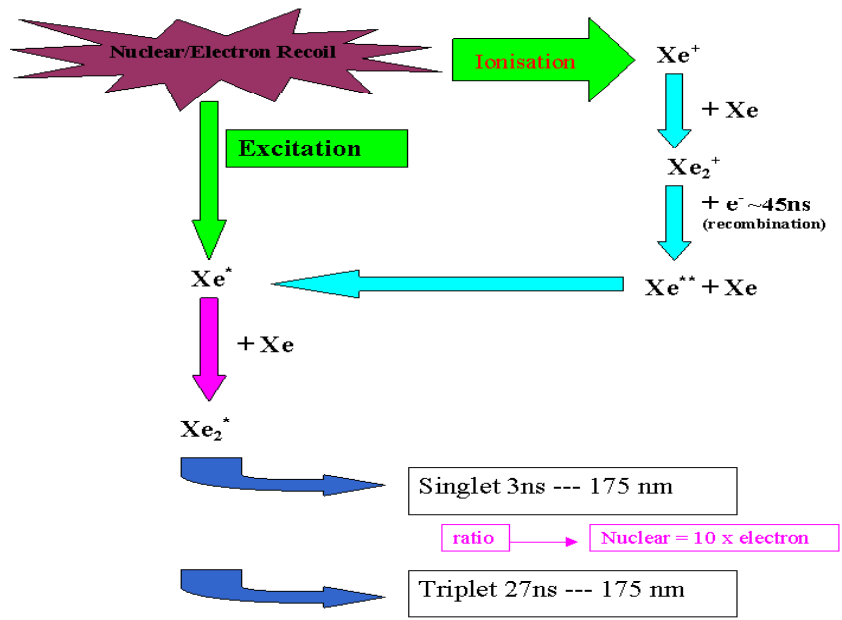

Figure 2. Basic mechanism for the signal and background detection in liquid Xe.

---

**TABLE 3.**

# Liquid Xenon as a WIMP Detector

1. Large mass available - up to tons.

- Atomic mass: 131.29
- Density: 3.057 gm/cm$^3$
- $W_i$ value (eV/pair) 15.6 eV
- No long-lived isotopes of xenon

2. Drift velocity: 1.7 mm/$\mu$.s @ 250V/cm field

                  X Scintillation wave length: 175 nm

- Decay time: 2 ns $\rightarrow$ 27 ns

3. Light yield > NaI, but intrinsic scintillator (no doping)

$\Rightarrow$ Excimer process very well understood

$\Rightarrow$ First excimer laser was liquid xenon in 1970!

---

### (b) CDMS II at Soudan

The CDMS II detector follows a long line of research and development on cryogenic methods. The current detector will be seven kilograms. Figure 6 shows a recent picture of the detector. This detector and Edelweiss I will likely have the lowest event threshold of any of the detectors illustrated in table 2.

### 4. Future One-Ton Detectors: ZEPLIN IV Development

To consider a 1-ton detector (ZEPLIN IV) we adopt the design principle of the ZEPLIN II detector under construction now. As we complete and operate ZEPLIN II, adjustments to the design of ZEPLIN IV will be carried out by this team [8]. The basic concept of ZEPLIN IV is shown in Figure 6 (from H. Wang design).

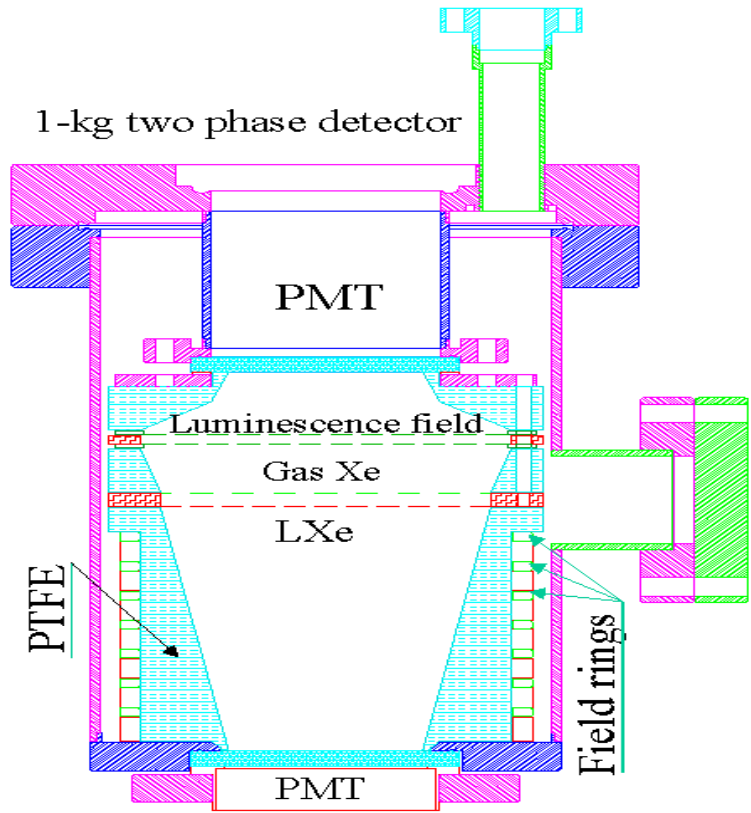

Fig. 3. UCLA/Torino Study Detector. Electroluminescence in gas (principle of a two-phase, 1-kg detector [4].

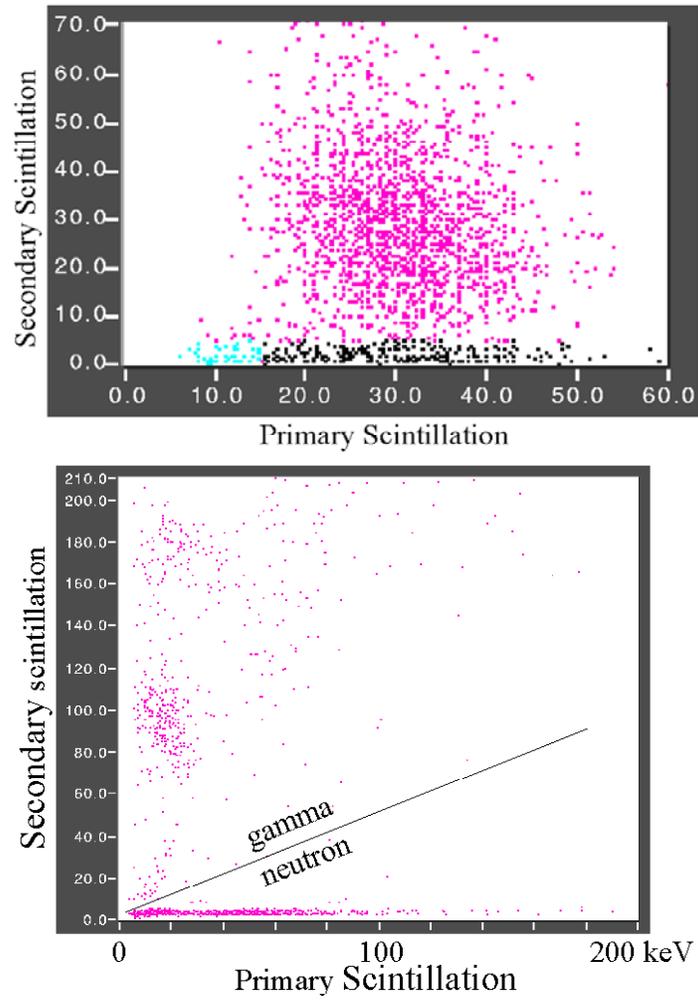

Fig. 4. Discrimination plots for (a) a Liquid Xenon Detector and (b) a Two Phase Liquid Xenon Detector (UCLA/Torino group) [4].

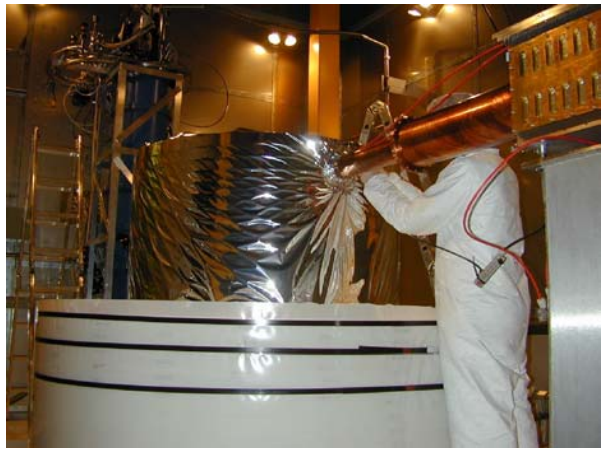

Fig. 5. The CDMS II Detector being assembled at Soudan Underground Laboratory. [1]

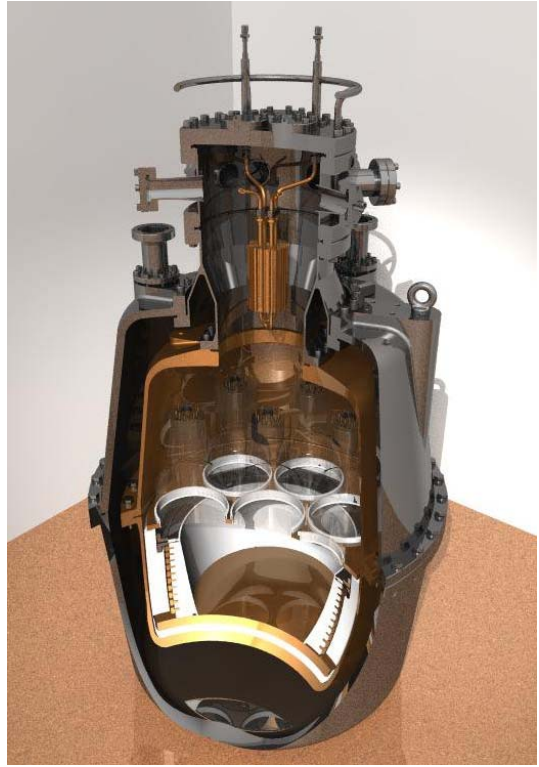

Fig. 6. Drawing of the ZEPLIN II Detector [6].

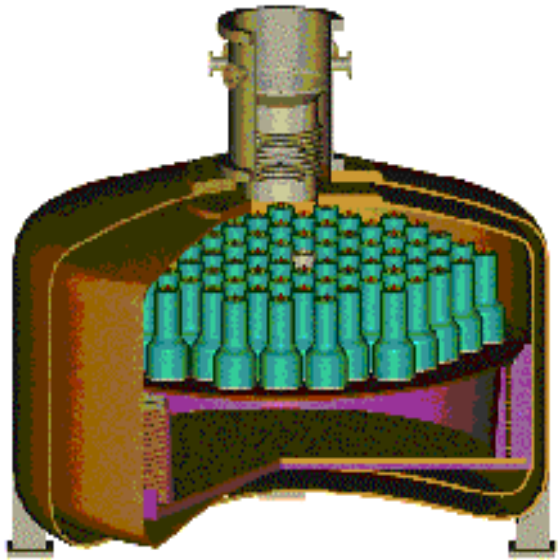

Fig. 7. A schematic of a one ton ZEPLIN IV Detector.

## 6. Current estimates for the SUSY dark matter signal level

At this meeting (SUGRA 20) new estimates for the cross-section for SUSY-WIMPs were given (see the proceedings). We follow the result of P. Nath and show a graph from his work in Figure 8 [2]. We also plot the sensitivity of ZEPLIN II and IV (as well as CDMS II and GENIUS) in Figure 8. It is clear that a one ton detector is needed to cover a significant part of the range of predictions. Nevertheless detectors like CDMS II, Edelweiss and ZEPLIN II may find the evidence for SUSY Dark Matter.

We thank P. Nath, H. Wang and members of the UKDMC-ZEPLIN program for discussions.

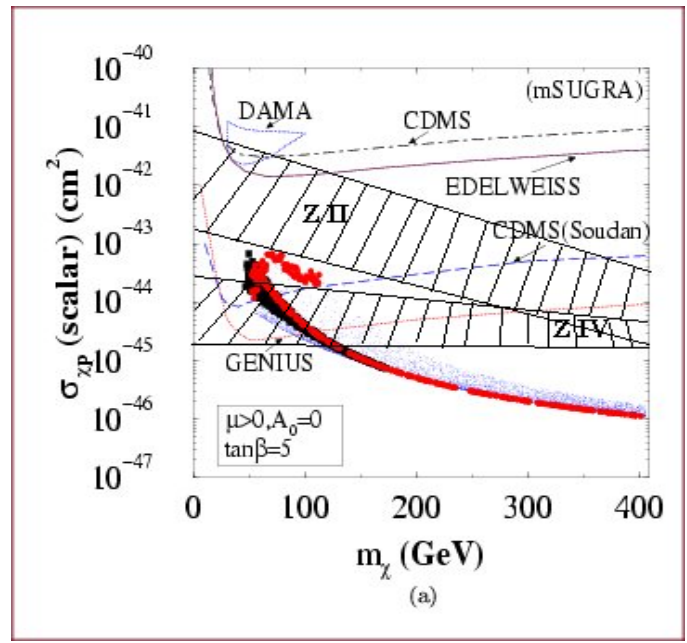

Figure 8. Expected cross-sections for SUSY WIMPs in SUGRA models (P. Nath) and some recent search results. The range of sensitivity for ZEPLIN II and IV are shown on the modified plot [2].